\documentclass[twocolumn,showpacs,preprintnumbers,amsmath,amssymb]{revtex4}

\usepackage{graphicx}
\usepackage{dcolumn}
\usepackage{bm}

\begin{document}

\title{Cosmological Evolution of Interacting Phantom
 Energy with Dark Matter}

\author{Zong-Kuan Guo}
\email{guozk@itp.ac.cn}
\author{Rong-Gen Cai}
\affiliation{
Institute of Theoretical Physics, Chinese Academy of Sciences,
P.O. Box 2735, Beijing 100080, China}
\author{Yuan-Zhong Zhang}
\affiliation{
CCAST (World Lab.), P.O. Box 8730, Beijing 100080\\
Institute of Theoretical Physics, Chinese Academy of Sciences,
P.O. Box 2735, Beijing 100080, China}

\date{\today}

\begin{abstract}
We investigate the cosmological evolution of an interacting phantom
energy model in which the phantom field has interaction with the
dark matter.
We discuss the existence and stability of scaling solutions for two
types of specific interactions. 
One is motivated by the conformal transformation in string theory
and the other is motivated by analogy with dissipation.
In the former case, there exist no scaling solutions.
In the latter case, there exist stable scaling solutions, which may
give a phenomenological solution of the coincidence problem.
Furthermore, the universe either accelerates forever or ends with
a singularity, which is determined by not only the model parameters
but also the initial velocity of the phantom field.
\end{abstract}

\maketitle

\section{Introduction}

Scalar field plays an important role in modern cosmology. The dark
energy can be attributed to the dynamics of a scalar field, for instance
quintessence~\cite{RP, ZWS}, which convincingly realize the present
accelerated expansion by using late-time attractor solutions, in which
the scalar field mimics the perfect fluid in a wide range of parameters.
Much attention has been drawn to the case of exponential potentials.
The exponential potentials allow the possible existence of scaling
solutions in which the scalar field energy density tracks that of the
perfect fluid (so that at late times neither component can be negligible).
In particular, a phase-plane analysis of the spatially flat FRW models
showed that these solutions are the unique late-time attractors
whenever they exist~\cite{CLW, ZKG}. Moreover,
exponential potentials appear naturally in the low dimensional models of
string/M-theory~\cite{CHN}.

The recent SNe data seem to favor the dark energy with the present
equation of state $w < -1$~\cite{ASSS}. To obtain $w < -1$,
phantom field with a negative kinetic term may be a simplest
implementing and can be regarded as one of interesting possibilities
describing dark energy~\cite{RRC}.
The physical background for phantom type of matter with strongly
negative pressure would be found in string theory~\cite{MGK}.
Phantom field may also arise in higher-order theories of
gravity~\cite{MDP}, Brans-Dicke and non-minimally coupled scalar
field theories~\cite{DFT}. The cosmological models which allow for
phantom matter appear naturally in k-essence models~\cite{COY}.
In spite of the fact that the field theory of phantom fields encounters
the problem of stability which one could try to bypass by assuming
them to be effective fields~\cite{GWG}, it is nevertheless interesting to
study their cosmological implication. Recently, there have been many
relevant studies on this topic~\cite{SW}.

The physical properties of phantom energy are rather weird, as they
include violation of the dominant energy condition and increasing
energy density with the expansion of the universe.
The latter ultimately leads to an unwanted future singularity
called big rip. This singularity is characterized by the
divergence of the scale factor in a finite time in future~\cite{CKW}.
To avoid the cosmic doomsday, some phantom field models were
proposed~\cite{CHT}. It requires a special class of phantom field
potentials with a local maximum. Moreover, the energy density of
the phantom field increases with time, while the energy density of
the matter fluid decreases as the universe expands. Why are the
energy density of dark matter and the phantom energy density of the
same order just at the present epoch? This coincidence problem
becomes more difficult to solve in the phantom model. Therefore,
from this point of view the cosmological \emph{scaling solution}
would be desirable for the history of the universe.
Throughout this paper we use ``scaling solution" as a meaning
that the energy densities of the phantom field and the dark matter
are proportional. However, as shown in Ref.~\cite{GPZZ},
there exist no scaling solutions because the phantom energy
increases while the matter energy decreases with time. But in the
presence of the suitable interaction this case could be realized
easily in Ref.~\cite{GZ}. In this paper we investigate the stability
and existence of scaling solutions in the scenario of interacting
phantom energy with dark matter.
We consider two phenomenological models. One is motivated
by the conformal transformation from the Jordan frame to Einstein
frame in string theory and the other is motivated by analogy with
dissipation. In the former model there exist no scaling solutions.
However, in the latter model a phase-plane analysis shows that there
exist two kinds of stable scaling solutions, which lead to two different
fates of the universe. The universe either accelerates forever or ends
with a big rip, which is determined by not only the model parameters
but also the initial velocity of the phantom field.
We will also discuss the physical consequences of these results.

\section{Interacting Phantom Energy with Dark Matter}

Let us consider a universe model where both the phantom field
$\phi$ and the dark matter $\rho_m$ are present. The Friedmann
equation in a spatially flat FRW metric can be written as
\begin{equation}
\label{FE}
H^2=\frac{\kappa ^2}{3}(\rho_p+\rho_m),
\end{equation}
where $\kappa^2 \equiv 8\pi G_N$ is the gravitational coupling
and the energy density and pressure, $\rho_p$ and $P_p$, of the
homogeneous phantom field $\phi$ are given by
\begin{eqnarray}
\rho_p &=& -\frac{1}{2}\dot{\phi}^2+V(\phi), \\
P_p   &=& -\frac{1}{2}\dot{\phi}^2-V(\phi),
\end{eqnarray}
respectively, in which $V(\phi)$ is the phantom field potential.
We postulate that the two components, $\rho_p$ and $\rho_m$,
interact through the interaction term $Q$ according to
\begin{eqnarray}
\label{EE1}
\dot{\rho}_m + 3H (\rho_m+P_m) &=& Q, \\
\dot{\rho}_p + 3H (\rho_p+P_p) &=& -Q.
\label{EE2}
\end{eqnarray}

Suppose the dark matter possesses the equation of state $P_m=0$.
The dynamics of the phantom field with an exponential potential
\begin{equation}
\label{EPV}
V(\phi)=V_0\exp (-\lambda \kappa \phi)
\end{equation}
has been analysed in Ref.~\cite{GPZZ}. We assume the dimensionless
constant $\lambda$ is positive since we can make it positive through
$\phi \to -\phi$ if $\lambda < 0$. Here we generalize the analysis
to the case in which the phantom field has interaction with the dark
matter. As we will see, this gives rise to some interesting novel features.
Following Ref.~\cite{CLW}, we define the following
dimensionless variables
\begin{equation}
x \equiv \frac{\kappa \dot{\phi}}{\sqrt{6}H}\,, \qquad
y \equiv \frac{\kappa \sqrt{V}}{\sqrt{3}H}\,, \qquad
z \equiv \frac{\kappa \sqrt{\rho_m}}{\sqrt{3}H}\,.
\end{equation}
Notice that $x^2$, $y^2$ and $z^2$ give the fraction of total energy
density carried by the field kinetic energy, the field potential energy
and the dark matter, respectively. Thus the fractional densities of
$\rho_p$ and $\rho_m$ can be written as $\Omega_p=-x^2 + y^2$
and $\Omega_m=z^2$, respectively. The evolution equations (\ref{EE1})
and (\ref{EE2}) can be written as the following
set of equations:
\begin{eqnarray}
\label{SE1}
x' &=& -3x\left(1+x^2-\frac{1}{2}z^2\right)+
 \frac{\kappa}{\sqrt{6}}\frac{Q}{H^2\dot{\phi}}-
 \frac{3}{\sqrt{6}}\lambda y^2, \\  
y' &=& -3y\left(x^2+ \frac{\sqrt{6}}{6}\lambda x-\frac{1}{2}z^2 \right), \\
z' &=& -3z\left(\frac{1}{2}+x^2-\frac{1}{2}z^2\right)+
 \frac{\kappa}{2\sqrt{3}}\frac{Q}{H^2\sqrt{\rho_m}}\,,
\label{SE3}
\end{eqnarray}
where the prime denotes a derivative with respect to the logarithm of
the scale factor, $N \equiv \ln a$, and the Fridemann constraint
equation (\ref{FE}) becomes 
\begin{equation}
-x^2 + y^2 + z^2 =1.
\end{equation}
The critical points, where $x'=0$, $y'=0$ and $z'=0$, correspond to an
expanding universe with a scale factor $a(t)$ given by
$a \propto t^{2/[3(1+w_{\mathrm{eff}})]}$. The effective equation of state
for the total comic fluid is
\begin{equation}
w_{\mathrm{eff}}=-x^2-y^2.
\end{equation}

Interaction terms $Q$ have been discussed in the literature within the
context of inflation and reheating. In the conventional reheating model,
an interaction term $\Gamma_\phi \dot{\phi}^2$ dominates at the end
of inflation when the scalar field is oscillating about the minimum of
its potential. During this reheating phase the energy transferred from
the scalar field is completely converted into the matter.
Within the context of exponential potentials, an interaction term of the
form $Q=-c\,\kappa\,\rho_m \,\dot{\phi}$ was considered in
Ref.~\cite{WETT,LAM}. It was shown that the matter scaling solutions
were stable solutions and the age of the universe is older when such
the interaction term is included. Certain string theories in which the
energy sources are separately conserved in the Jordan frame naturally
lead to interaction terms in the Einstein frame; scalar tensor theory with
matter terms may yield the same results~\cite{WETT,LAM}.
An interaction term of the form $Q=3\,c\,H(\rho_p+\rho_m)$ was
proposed to look for a dynamical solution to the coincidence
problem~\cite{ZPC}. Such the interaction term might be motivated by
analogy with dissipation. For example, a fluid with bulk viscosity may
give rise to a term of this form in the conservation equation.
Without an interaction term, it was shown that the dark matter could
not track the phantom energy and would be quickly driven to
zero~\cite{GPZZ}. It is of interest to study the cosmological
consequences of the above two types of interactions in the phantom
energy model.

\section{Model I}

Let us first consider the following interaction~\cite{WETT,LAM,TS}
\begin{equation}
\label{IT1}
Q=-c\,\kappa\,\rho_m \,\dot{\phi},
\end{equation}
where $c$ is a dimensionless parameter. Such a coupling arises for
instance in string theory~\cite{WETT}, or after a conformal transformation
of Brans-Dicke theory~\cite{LAM}.
The evolution equations (\ref{SE1}-\ref{SE3}) can then be written as
an autonomous system:
\begin{eqnarray}
\label{AS1}
x' &=& -3x\left(1+x^2-\frac{1}{2}z^2\right)-\frac{3}{\sqrt{6}}c\,z^2-
 \frac{3}{\sqrt{6}}\lambda y^2, \\  
y' &=& -3y\left(x^2+ \frac{\sqrt{6}}{6}\lambda x-\frac{1}{2}z^2 \right), \\
z' &=& -3z\left(\frac{1}{2}+x^2-\frac{1}{2}z^2+\frac{\sqrt{6}}{6}c\,x\right),
\label{AS3}
\end{eqnarray}
which has three critical points as follows.

\emph{Point A}:
\begin{equation}
x_A=-\frac{\sqrt{6}}{3}c, \qquad y_A=0,
 \qquad z_A=\sqrt{1+\frac{2}{3}c^2}\,.
\end{equation}
This solution is physically meaningless since $\Omega_m > 1$
if $c \ne 0$. 

\emph{Point B}:
\begin{equation}
x_B=-\frac{\sqrt{6}}{6}\lambda,
 \qquad y_B=\sqrt{1+\frac{\lambda^2}{6}}, \qquad z_B=0.
\end{equation}
This critical point corresponds to the phantom-dominated solution
$\Omega_p=1$, which always exists for any $\lambda$ and $c$.
The effective equation of state, $w_{\mathrm{eff}}=-1-\lambda^2/3$,
depends on the slope of the potential. To find out under what
condition this fixed
point is a stable solution, we study the behavior of small deviations
from the solution. The linearization of system (\ref{AS1})-(\ref{AS3})
about this fixed point yields two eigenvalues
$m_1=-(3+\lambda^2/2)$ and $m_2=-(\lambda^2+3-c\,\lambda)$.
Thus the phantom-dominated solution is stable for
$c \le \lambda+3/\lambda$.

\emph{Point C}:
\begin{eqnarray}
x_C &=& \frac{3}{\sqrt{6}(\lambda-c)}, \qquad
y_C^2 = 1-\frac{\lambda}{\lambda-c}-\frac{3}{2(\lambda-c)^2},
 \nonumber \\
z_C^2 &=& \frac{\lambda}{\lambda-c}+\frac{3}{(\lambda-c)^2}.
\end{eqnarray}
This fixed point corresponds to the phantom-fluid scaling solution,
which exists for
\begin{displaymath}
c \le \frac{\lambda-\sqrt{\lambda^2+12}}{2} \quad \textrm{or} \quad
\frac{\lambda+\sqrt{\lambda^2+12}}{2} \le c \le \lambda+\frac{3}{\lambda}.
\end{displaymath}
Substituting linear perturbations $x \to x_C+\delta x$,
$y \to y_C+\delta y$, and $z \to z_C+\delta z$
about the critical point into the system of
equations (\ref{AS1})-(\ref{AS3}), to first-order in the perturbations,
gives the following two independent evolution equations of the linear
perturbations:
\begin{eqnarray}
\delta x' &=& -\frac{3}{2}\left(1+3x_C^2+y_C^2+
 \frac{4}{\sqrt{6}}c\,x_C^{}\right) \delta x \nonumber \\
        & & -\left[3x_C^{}\,y_C^{}+
 \frac{6}{\sqrt{6}}(\lambda-c)y_C^{}\right]\delta y,\\
\delta y' &=& -3y_C^{}\left(\frac{\sqrt{6}}{6}\lambda+
 x_C^{}\right) \delta x- 3y_C^2 \delta y.
\end{eqnarray}
The two eigenvalues of the coefficient matrix of the above equations
determine the stability of the critical point. We find that the solution
is unstable.

\begin{figure}
\begin{center}
\includegraphics[scale=0.8]{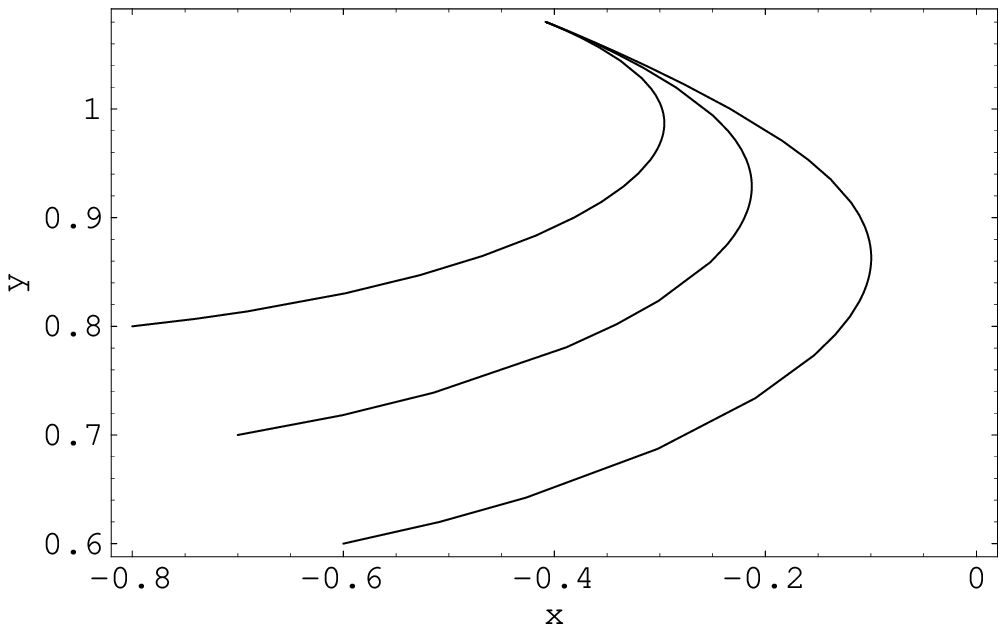}
\end{center}
\caption{The convergence of different initial conditions to the attractor
solution in the ($x$, $y$) phase space for the model I with $c=-2$
and $\lambda=1$.}
\end{figure}

\begin{figure}
\begin{center}
\includegraphics[scale=0.8]{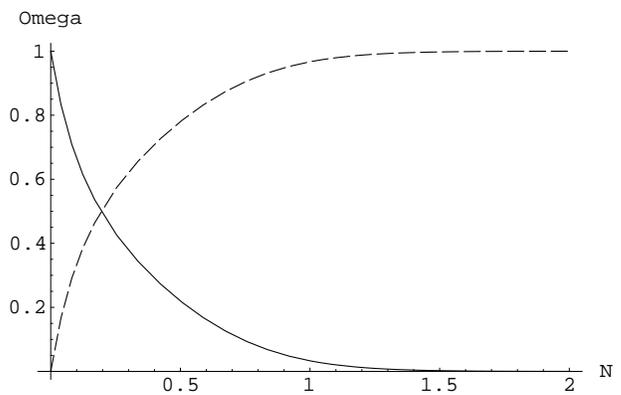}
\end{center}
\caption{The evolution of the fractional densities of the phantom field
(the dashed curve) and the dark matter (the solid curve) for the case
in Fig.1.}
\end{figure}

In the cosmological model with the interaction (\ref{IT1}) between
phantom field and dark matter, the phantom-dominated solution is the
only attractor solution in the parameter space, $c\le\lambda+3/\lambda$.
In Figs.1-4, we plot the numerical results. Comparing Fig.2 to Fig.4,
we see that the phantom energy more quickly dominates the universe
when the parameter $c$ decreases. In Fig.1 and Fig.3, the trajectories
converge at the same fixed point, which is only determined by the
parameter $\lambda$.
Hence energy transfer whether from the phantom field to the dark
matter or vice versa yields the similar cosmological consequences.
The stable critical point $B$ with $x_B < 0$
indicates that the phantom field climbs up the exponent
potential. The energy density of the phantom field increases as the
universe expands, which
leads to unwanted future singularity, and therefore the coincidence
problem becomes more difficult. In the next section, we will
investigate a phenomenal model, in which the cosmic doomsday is
avoided and the coincidence problem may be alleviated.

\begin{figure}
\begin{center}
\includegraphics[scale=0.8]{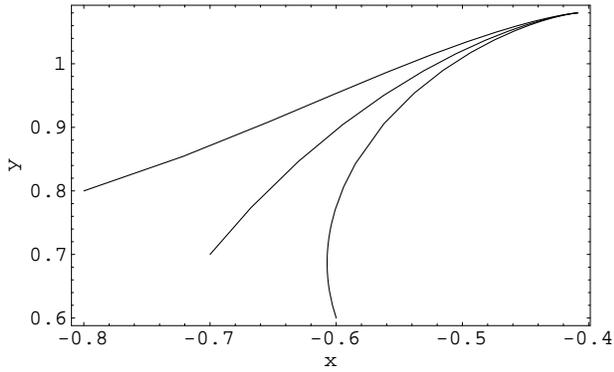}
\end{center}
\caption{The convergence of different initial conditions to the attractor
solution in the ($x$, $y$) phase space for the model I. We choose
$c=1$ and $\lambda=1$.}
\end{figure}

\begin{figure}
\begin{center}
\includegraphics[scale=0.8]{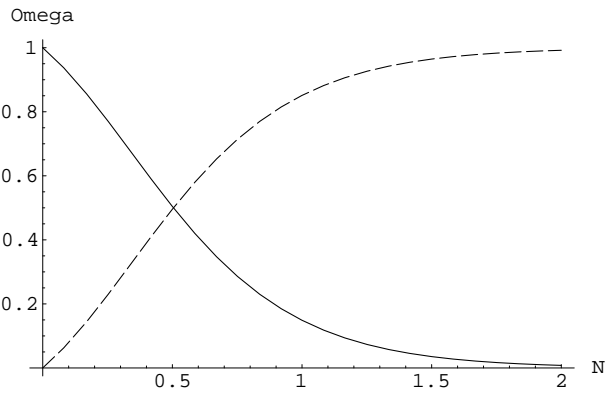}
\end{center}
\caption{The evolution of the fractional densities of the phantom field
(the dashed curve) and the dark matter (the solid curve) for the case
in Fig.3.}
\end{figure}

\section{Model II}

Now let us consider the specific interaction~\cite{ZPC,GZ,CW}
\begin{equation}
\label{IT2}
Q=3\,c\,H(\rho_p+\rho_m),
\end{equation}
where $c$ is a dimensionless parameter denoting the transfer strength.
This type of interaction has been proposed to look for a dynamical
solution to the coincidence problem in Ref.~\cite{ZPC}.
Then the equation system (\ref{SE1}-\ref{SE3}) can be written as
an autonomous system:
\begin{eqnarray}
x' &=& -3x\left(1+x^2-\frac{1}{2}z^2-\frac{1}{2}c\,x^{-2}\right)-
 \frac{3}{\sqrt{6}}\lambda y^2, \\  
y' &=& -3y\left(x^2+ \frac{\sqrt{6}}{6}\lambda x-\frac{1}{2}z^2 \right), \\
z' &=& -3z\left(\frac{1}{2}+x^2-\frac{1}{2}z^2-\frac{1}{2}c\,z^{-2}\right),
\end{eqnarray}
which has four critical points.

\emph{Point A}:
\begin{eqnarray}
x_A^2 &=& \frac{1}{2}(\sqrt{1+4c}-1),
 \quad y_A=0, \nonumber \\
z_A^2 &=& \frac{1}{2}(\sqrt{1+4c}+1).
\end{eqnarray}
This solution is physically meaningless since $\Omega_p < 0$.

\emph{Points B,C,D}:
The other three critical points are solutions of the following set of
equations:
\begin{eqnarray}
f(x) &=& c, \\
y^2 &=& -x^2-\frac{\sqrt{6}}{3}\lambda x + 1, \\
z^2 &=& 2x^2+\frac{\sqrt{6}}{3}\lambda x,
\end{eqnarray}
where we have defined a cubic function
\begin{equation}
f(x) \equiv x\left(2x+\frac{\sqrt{6}}{3}\lambda\right)
 \left(1-\frac{\sqrt{6}}{3}\lambda x \right).
\end{equation}
The critical point with $x_B < 0$, labelled by \emph{B}, exists for
\begin{displaymath}
0 < c \le f(\frac{-\lambda-\sqrt{\lambda^2+12}}{2\sqrt{6}}).
\end{displaymath}
There are two critical points with $x_{C,D}> 0$, one of which is
physically
meaningless, labelled by \emph{D}. The other point, labelled by
\emph{C}, exists for
\begin{displaymath}
0 < c \le
 \min\{f(\frac{-\lambda+\sqrt{\lambda^2+6}}{\sqrt{6}}),
 f(\frac{-\lambda+\sqrt{\lambda^2+12}}{2\sqrt{6}})\}.
\end{displaymath}
We see that the point \emph{B} corresponds to a climbing-up
phantom field, while the point \emph{C} corresponds to a rolling-down
phantom field. In order to study the stability of the two critical points,
we obtain the two independent evolution equations of the linear
perturbations
\begin{eqnarray}
\delta x' &=& -\frac{3}{2}\left(1+3x_{B,C}^2+y_{B,C}^2+
 \frac{c}{x_{B,C}^2}\right) \delta x \nonumber \\
        && -\left(3x_{B,C}^{}\,y_{B,C}^{}+
 \frac{6}{\sqrt{6}}\lambda y_{B,C}^{}\right) \delta y, \\
\delta y' &=& -3y_{B,C}^{}\left(\frac{\sqrt{6}}{6}\lambda+
 x_{B,C}^{}\right) \delta x- 3y_{B,C}^2\, \delta y.
\end{eqnarray}
The corresponding eigenvalues of the coefficient matrix of the above
equations indicate that the critical points \emph{B} and \emph{C} are
always the late-time stable attractor solutions if they exist. 

\begin{figure}
\begin{center}
\includegraphics[scale=0.8]{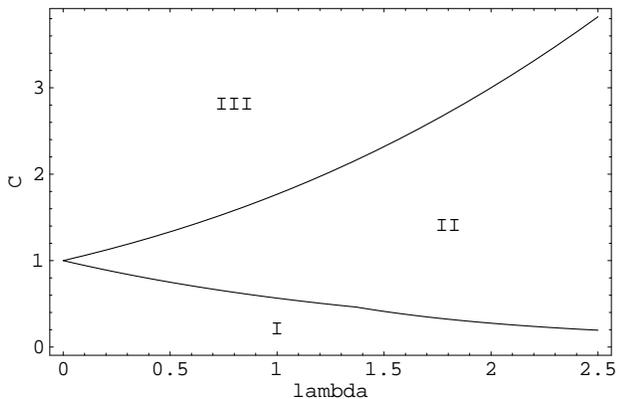}
\end{center}
\caption{Stability regions of the ($\lambda$, $c$) parameter space
for the model II. 
In the regions I, either the climbing-up scaling solution or the
rolling-down scaling solution is the stable late-time attractor.
In the region II, the climbing-up solution is the stable late-time
attractor. The solutions are physically meaningless in the region III.}
\end{figure}

In the case of the interaction form (\ref{IT2}), there exist two kinds
of stable scaling solutions, the climbing-up scaling solution with
$x_B<0$ in Figs.6-8 and the rolling-down scaling solution
with $x_C>0$ in Figs.9-11. As shown in Fig.7 and Fig.10,
the universe evolves from the matter-dominated phase to the scaling
solution, which is characterized by a constant ratio of the energy
densities of the dark matter and the phantom field. This may provide
us with a phenomenological solution of the coincidence problem.
These results agree with those in Ref.~\cite{GZ}.
The different regions in the ($\gamma$, $c$) parameter space lead to
different qualitative evolutions. In the region II of th parameter space
in Fig.5, the critical point \emph{B} is a stable solution. However, in
the region I both the points \emph{B} and \emph{C} are stable. Which
one is the late-time stable attractor solution? The phantom field either
climbs up or rolls down the exponent potential, which is determined
by the initial velocity of the phantom field. If the phantom field initially
climbs up, the effective equation of state $w_{\mathrm{eff}}$ tends to
below $-1$ and realizes a transition from $w_{\mathrm{eff}}>-1$ to
$w_{\mathrm{eff}}<-1$ in Fig.8. Thus the universe ends with a big rip.
If the phantom field initially rolls down, the effective equation of state
$w_{\mathrm{eff}}$ tends to above $-1$ and realizes a transition from
$w_{\mathrm{eff}}<-1$ to $w_{\mathrm{eff}}>-1$ in Fig.11. In this
case the cosmic doomsday is avoided and the universe accelerates
forever.

\begin{figure}
\begin{center}
\includegraphics[scale=0.8]{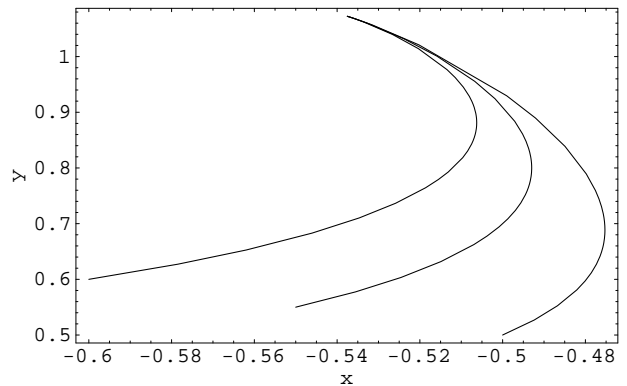}
\end{center}
\caption{The convergence of different initial conditions to the attractor
solution in the ($x$, $y$) phase space for the model II with $c=0.2$
and $\lambda=1$. We choose initial conditions with $x_0 < 0$.}
\end{figure}

\begin{figure}
\begin{center}
\includegraphics[scale=0.8]{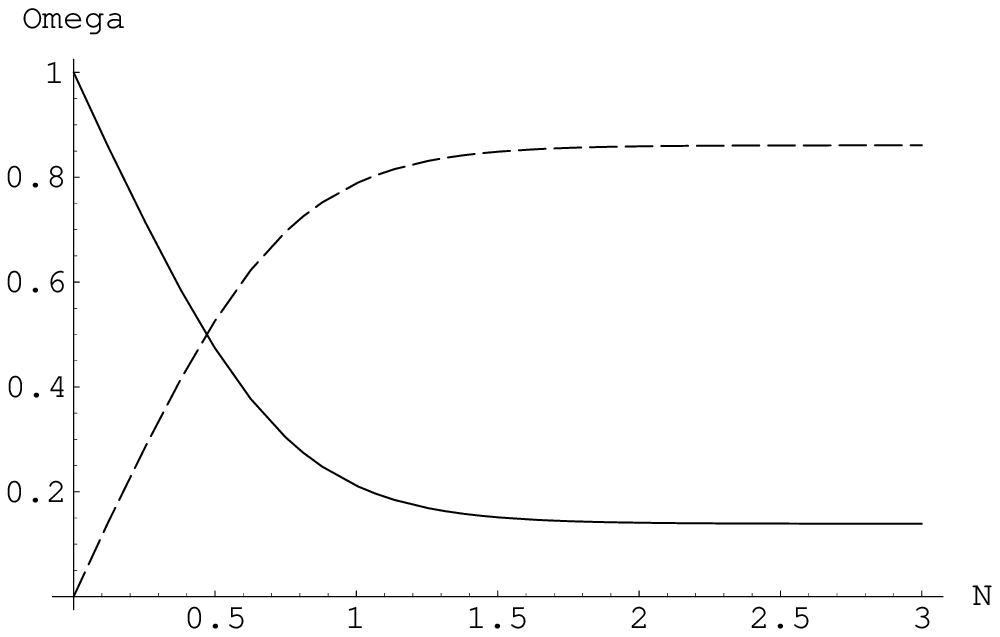}
\end{center}
\caption{The evolution of the fractional densities of the phantom field
(the dashed curve) and the dark matter (the solid curve) for the case
in Fig.6.}
\end{figure}

\begin{figure}
\begin{center}
\includegraphics[scale=0.8]{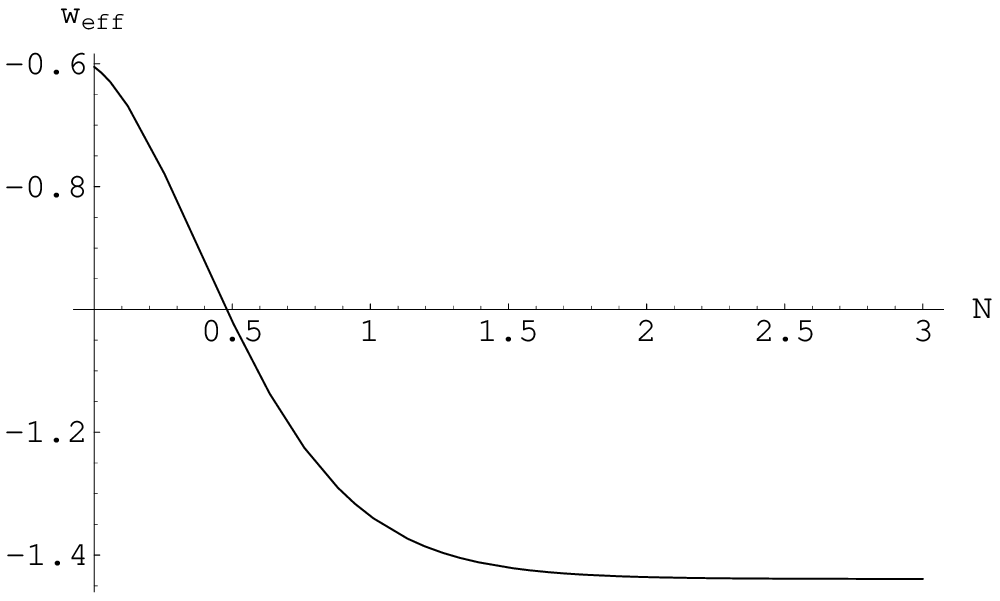}
\end{center}
\caption{The evolution of the effective equation of state for the case
in Fig.6.}
\end{figure}

\begin{figure}
\begin{center}
\includegraphics[scale=0.8]{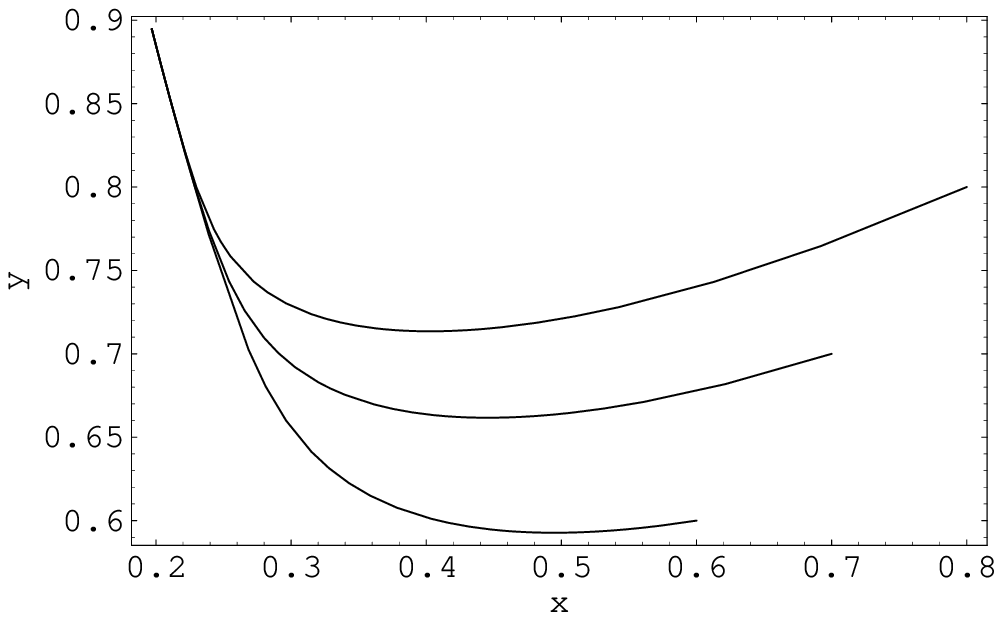}
\end{center}
\caption{The convergence of different initial conditions to the attractor
solution in the ($x$, $y$) phase space for the model II with $c=0.2$
and $\lambda=1$. We choose initial conditions with $x_0 > 0$.}
\end{figure}

\begin{figure}
\begin{center}
\includegraphics[scale=0.8]{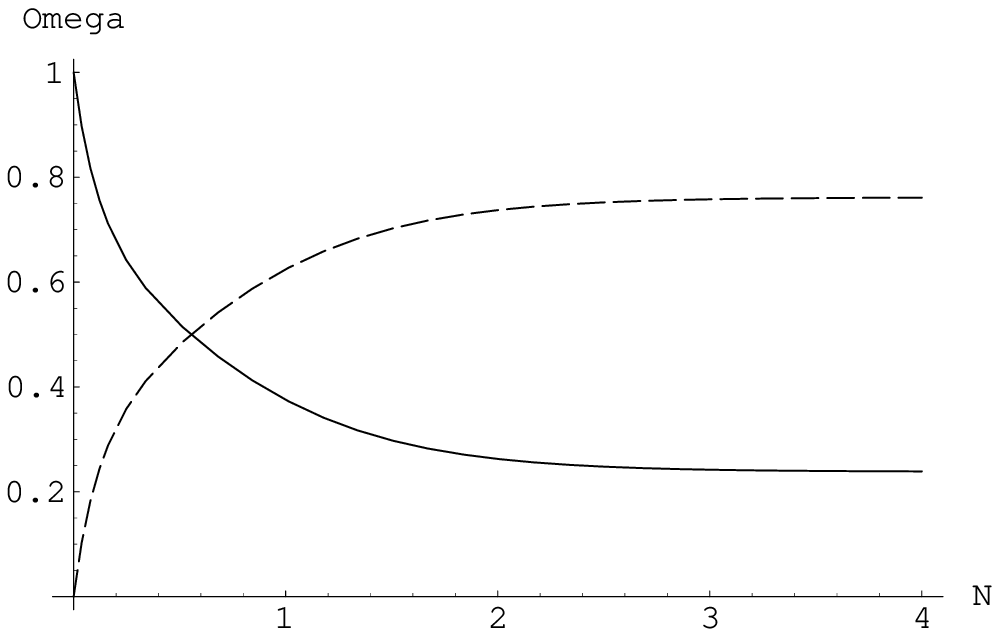}
\end{center}
\caption{The evolution of the fractional densities of the phantom field
(the dashed curve) and the dark matter (the solid curve) for the case
in Fig.9.}
\end{figure}

\begin{figure}
\begin{center}
\includegraphics[scale=0.8]{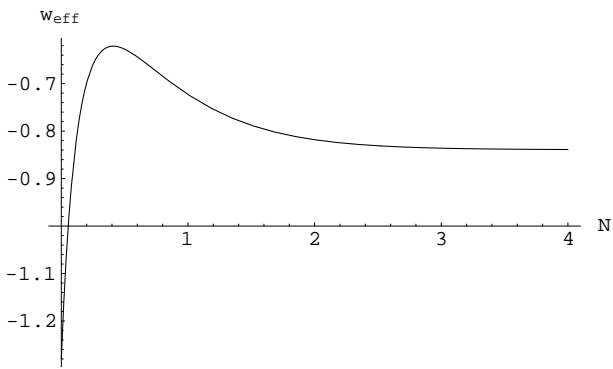}
\end{center}
\caption{The evolution of the effective equation of state for the case
in Fig.9.}
\end{figure}

\section{Conclusions and Discussions}

We have presented a phase-space analysis of the evolution for a
spatially flat FRW universe driven by an interacting mixture of dark
matter and phantom field with an exponent potential.
We have discussed the existence and stability of the cosmological
scaling solution for two types of interactions, namely
$Q=-c\,\kappa\,\rho_m \,\dot{\phi}$ motivated by the conformal
relationships between the Jordan and Einstein frame in string theory,
and $Q=3\,c\,H(\rho_p+\rho_m)$
motivated by analogy with dissipation.
In the former model, the phantom-dominated solution is the only
attractor solution when $c\le\lambda+3/\lambda$.
Energy transfer whether from the phantom field to the dark matter
(i.e. $c>0$) or vice versa (i.e. $c<0$) leads to similar behavior,
except that the phantom energy more quickly dominates the
universe in the latter case than the former case.
Since the phantom field climb up the exponent potential,
the energy density of the phantom field quickly increases as the
universe expands, which leads to an unwanted future singularity.

However, in the model II with the interaction (\ref{IT2}) between
phantom field and dark matter, there exist two kinds of stable
scaling solutions, the climbing-up scaling solution and
the rolling-down scaling solution. The existence
of a stable scaling solution requires a transfer of energy from the
phantom field to the dark matter. In this model the universe evolves
from a matter-dominated phase to a scaling solution, which is
characterized by a constant ratio of the energy densities of the dark
matter and the phantom field. This may provide us with a
phenomenological solution of the coincidence problem.
Furthermore, in the climbing-up case, the effective equation of state
$w_{\mathrm{eff}}$ tends to below $-1$, and then the universe
ends with a big rip. The effective equation of state may realize a
transition from $w_{\mathrm{eff}}>-1$ to $w_{\mathrm{eff}}<-1$.
In the rolling-down case, the effective equation of state
$w_{\mathrm{eff}}$ tends to above $-1$ and may cross $-1$.
In this case the cosmic doomsday is avoided and the universe
accelerates forever with a power-law form. What is the
ultimate fate of the universe? The universe either accelerates forever
or ends with a big rip, which is determined by not only the
model parameters but also the initial velocity of the phantom field
in the scenario of interacting phantom energy with dark matter.

\begin{acknowledgments}
This project was in part supported by National Basic Research
Program of China under Grant No.2003CB716300, 
by NNSFC under Grant No.90403032, No.10325525 and No.90403029,
and also by MSTC under Grant No.TG1999075401.
\end{acknowledgments}


\begin{thebibliography}{99}
\bibitem{RP}
 B.Ratra and P.J.E.Peebles, Phys.Rev. {\bf D37} (1988) 3406;
 C.Wetterich, Nucl.Phys. {\bf B302} (1988) 668.
\bibitem{ZWS}
 I.Zlatev, L.M.Wang and P.J.Steinhardt, Phys.Rev.Lett. {\bf 82} (1999) 896;
 P.J.Steinhardt, L.Wang and I.Zlatev, Phys.Rev. {\bf D59} (1999) 123504.
\bibitem{CLW}
 E.J.Copeland, A.R.Liddle and D.Wands, Phys.Rev. {\bf D57} (1998) 4686.
\bibitem{ZKG}
 Z.K.Guo, Y.S.Piao and Y.Z.Zhang, Phys.Lett. {\bf B568} (2003) 1;
 Z.K.Guo, Y.S.Piao, R.G.Cai and Y.Z.Zhang, Phys.Lett. {\bf B576} (2003) 12;
 Z.K.Guo and Y.Z.Zhang JCAP {\bf 0408} (2004) 010.
\bibitem{CHN}
 C.M.Chen, P.M.Ho, I.P.Neupane, N.Ohta and J.E.Wang, JHEP {\bf 0310} (2003) 058;
 M.N.R.Wohlfarth, Phys.Rev. {\bf D69} (2004) 066002;
 N.Ohta, hep-th/0411230.
\bibitem{ASSS}
 U.Alam, V.Sahni and A.A.Starobinsky, astro-ph/0403687;
 T.R.Choudhury and T.Padmanabhan, astro-ph/0311622;
 D.Huterer and A.Cooray, astro-ph/0404062;
 B.Feng, X.L.Wang and X.Zhang, astro-ph/0404224;
 Y.Gong, astro-ph/0405446.
\bibitem{RRC}
 R.R.Caldwell, Phys.Lett. {\bf B545} (2002) 23.
\bibitem{MGK}
 L.Mersini, M.Bastero-Gil and P.Kanti, Phys.Rev. {\bf D64} (2001) 043508;
 F.Piazza and S.Tsujikawa, JCAP 0407 (2004) 004.
\bibitem{MDP}
 M.D.Pollock, Phys.Lett. {\bf B215} (1988) 635;
 G.Calcagni, gr-qc/0410027.
\bibitem{DFT}
 D.F.Torres, Phys.Rev. {\bf D66} (2002) 043522;
 G.Esposito-Farese and D.Polarski, Phys.Rev. D63 (2001) 063504;
 E.Elizalde, S.Nojiri and S.D.Odintsov, Phys.Rev. {\bf D70} (2004) 043539.
\bibitem{COY}
 T.Chiba, T.Okabe and M.Yamaguchi, Phys.Rev. {\bf D62} (2000) 023511;
 J.M.Aguirregabiria, L.P.Chimento and R.Lazkoz, Phys.Rev. {\bf D70} (2004) 023509.
\bibitem{GWG}
 G.W.Gibbons, hep-th/0302199;
 V.K.Onemli and R.P.Woodard, Phys.Rev. {\bf D70} (2004) 107301;
 S.Nojiri and S.D.Odintsov, Phys.Lett. {\bf B562} (2003) 147;
 S.Nojiri and S.D.Odintsov, Phys.Lett. {\bf B565} (2003) 1.
\bibitem{SW}
 A.E.Schulz and M.White, Phys.Rev. {\bf D64} (2001) 043514;
 B.McInnes, astro-ph/0210321;
 M.P.Dabrowski, T.Stachowiak and M.Szydlowski, Phys.Rev. {\bf D68} (2003) 103519;
 Y.S.Piao and E.Zhou, Phys.Rev. {\bf D68} (2003) 083515;
 V.B.Johri, Phys.Rev. {\bf D70} (2004) 041303;
 Y.H.Wei and Y.Tian, Class.Quant.Grav. {\bf 21} (2004) 5347;
 Y.S.Piao and Y.Z.Zhang, astro-ph/0401231;
 J.Lima and J.S.Alcaniz, astro-ph/0402265;
 M.Bouhmadi-Lopez and J.J.Madrid, astro-ph/0404540;
 B.Feng, M.Li, Y.S.Piao and X.Zhang, astro-ph/0407432.
\bibitem{CKW}
 R.R.Caldwell, M.Kamionkowski and N.N.Weinberg, Phys.Rev.Lett. {\bf 91} (2003) 071301;
 P.F.Gonzalez-Diaz, Phys.Rev. D68 (2003) 021303;
 M.Sami and A.Toporensky, Mod.Phys.Lett. {\bf A19} (2004) 1509;
 H.Stefancic, Phys.Lett. {\bf B595} (2004) 9;
 L.P.Chimento and R.Lazkoz, gr-qc/0405020;
 S.Nesseris and L.Perivolaropoulos, astro-ph/0410309.
\bibitem{CHT}
 S.M.Carroll, M.Hoffman and M.Trodden, Phys.Rev. {\bf D68} (2003) 023509;
 P.Singh, M.Sami and N.Dadhich, Phys.Rev. {\bf D68} (2003) 023522;
 Z.K.Guo, Y.S.Piao and Y.Z.Zhang, Phys.Lett. {\bf B594} (2004) 247;
 I.Y.Arefeva, A.S.Koshelev and S.Y.Vernov, astro-ph/0412619.
\bibitem{GPZZ}
 Z.K.Guo, Y.S.Piao, X.Zhang and Y.Z.Zhang, astro-ph/0410654.
\bibitem{GZ}
 Z.K.Guo and Y.Z.Zhang, astro-ph/0411524.
\bibitem{WETT}
 C.Wetterich, Astron.Astrophys. {\bf 301} (1995) 321.
\bibitem{LAM}
 L.Amendola, Phys.Rev. {\bf D60} (1999) 043501.
\bibitem{ZPC}
 W.Zimdahl, D.Pavon and L.P.Chimento, Phys.Lett. {\bf B521} (2001) 133;
 L.P.Chimento, A.S.Jakubi, D.Pavon and W.Zimdahl, Phys.Rev. {\bf D67} (2003) 083513.
\bibitem{TS}
 A.P.Billyard and A.A.Coley, Phys.Rev. {\bf D61} (2000) 083503;
 S.Tsujikawa and M.Sami, Phys.Lett. {\bf B603} (2004) 113;
 H.Wei and R.G.Cai, hep-th/0412045.
\bibitem{CW}
 R.G.Cai and A.Wang, hep-th/0411025.
\end{thebibliography}
\end{document}